\documentclass[aps,preprint]{revtex4}
\usepackage{epsfig}
\usepackage{amsmath}
\usepackage{epsfig}

\begin{document}
\title
{Real discrete spectrum  in the complex non-PT-symmetric Scarf II potential}    
\author{Zafar Ahmed$^1$ and Joseph Amal Nathan$^2$\\
$~^1$Nuclear Physics Division, Bhabha Atomic Research Centre, Mumbai 400 085, India \\
$~^2$Reactor Physics Design Division, Bhabha Atomic Research Centre, Mumbai 400085, India}
\email{1:zahmed@barc.gov.in, 2:josephan@barc.gov.in}
\affiliation{}
\date{\today}
\begin{abstract}
Hitherto, it is well known that complex PT-symmetric
Scarf II has real discrete spectrum in the parametric domain of unbroken PT-symmetry.
We reveal new interesting complex, non-PT-symmetric parametric domains of this versatile potential, $V(x)$, where the spectrum is again  discrete and real. Showing that the Hamiltonian, $p^2/2\mu+V(x)$, in the new cases is pseudo-Hermitian could be challenging, if possible.
\end{abstract}
\maketitle

The Scarf II potential 
\begin{equation}
V(x)=P~ \mbox{sech}^2 x + Q~ \mbox{sech} x~ \tanh x,
\end{equation}
where
\begin{equation}
P=(b^2-a^2-a), Q=b(2a+1), a,b \in {\cal R}
\end{equation}
has been  one [1-5] of the most commonly used exactly
solvable potentials for both bound and scattering states. It admits beautiful expressions for discrete eigenvalues [1-4] and reflection/transmission amplitudes: $r(k)$, and $t(k)$ [5] that have been serving like a  facility. Consequently, this potential has contributed immensely in super-symmetric quantum mechanics, group theoretic and several other quantum mechanical studies.

After the advent of complex PT-symmetric quantum mechanics [6], for about one and a half decades, the Scarf II (1)  potential has been complexified time and again to reveal interesting features of the new version of quantum mechanics. Surprisingly, the parametrization of $P$ and $Q$ in various different ways has been unraveling  many interesting results [7-11,13,14,21-26].

By setting
\begin{equation}
P=-(b^2+a^2+a) ~ \mbox{and} ~ Q=ib(2a+1), a,b \in {\cal R}
\end{equation}
first [7] it was shown that a real Hermitian
and a complex PT-symmetric  potential can share an  identical energy spectrum. This work inspired the complexification of other [8] exactly solvable potentials 
as new solvable models of complex PT-symmetric potentials. Later,
using the parametrization (3), a new feature of complex PT-symmetry was revealed with regard to two branches [9] of real discrete energy spectrum of (1,3). These two branches were later discussed in terms of quasi-Parity [10]. A simple parametrization: $P=-V_1,V_1>0$ and $Q=iV_2$ [11] turned out to be most interesting when both unbroken (broken) domains of 
PT-symmetry were brought out explicitly and analytically as
$|V_2| \le (>) V_c$. Here, $V_c= V_1+1/4 ~(2\mu= 1= \hbar^2)$ which can now be called  the exceptional point (EP) of the complex potential. Such a phase-transition of eigenvalues has inspired experimental verifications [12] and further investigations [13,14]. For the parametrization (3), the reflection and transmission amplitudes have been re-derived [15] for (1) with (3) and a small yet important correction to those available in Ref. [5] has been suggested. 

More fruitfully, PT-symmetry has inspired investigations of
one-dimensional scattering from non-Hermitian potentials.
Though an abundant use of non-Hermitian central potential in three-dimensions in nuclear and condensed matter physics  existed before. The one-dimensional scattering started off late with non-reciprocity [16] of reflection probability with regard to the injection from left/right. Spectral singularity [17], coherent perfect absorption without [18] and with [19] lasing and transparency [20,21]  etc. are the novel phenomena which have been discovered. In this regard too, the complex Scarf II potential has not only provided simple  explicit demonstration of such phenomena [21-26] but also it is found to possesses the power to predict [22,27] some general features of scattering from complex non-Hermitian potentials.  
 
In this Letter, we reveal three non-PT-symmetric domains of complexified Scarf (II) potential which give rise to real discrete spectrum. 
 
So far, when  non-PT-symmetric [28-30] complex Hamiltonians have real discrete spectrum, they could be shown [29,30] to be pseudo-Hermitian: $\eta H \eta^{-1}=H^\dagger$ [31]. It must be remarked that before the advent of complex PT-symmetry [6] and pseudo-Hermiticity [31], in the studies of (de)localization of phase-transitions in non-Hermitian Hamiltonian, $H_{H-N}=(p+ig)^2/2\mu+V(x)$, the discrete eigenvalues were found to be real for $g <g_c$. This Hamiltonian has been later found [30] to be pseudo-Hermitian under the gauge-like transformation wherein $\eta=e^{-2gx}$.

However, we feel, that showing $H=p^2/(2\mu)+V(x)$, where $V(x)$ (to be proposed  here in sequel in  Eqs. (13,15,17)), as pseudo-Hermitian could be challenging, if possible. For an interesting technique of constructing pseudo-Hermitian Hamiltonians with real discrete spectrum see Ref. [32].

In the following, we propose to solve Scarf II, yet again however for first time as a non-Hermitian, non-PT-symmetric potential for  bound states. We use the parametrization for $P$ and $Q$ as 
\begin{equation}
P=-(B^2+A^2+A), Q= iB(2A+1).
\end{equation}
Here in contrast to (3), $A$ or $B$  may be complex. We re-analyze $V(x)$ (1) with (4) in Schr{\"o}dinger equation 
\begin{equation}
\frac  {d^2 \Psi(x)}{dx^2}+ \frac{2\mu}{\hbar^2}[E-V(x)]\psi(x)=0.
\end{equation}
At the out set, let us claim that the reality of eigenvalues when
$A$ or $B$ in (4) become non-real such that $V(x)$  is non-PT-symmetric is both new and surprising. Therefore,  
here, for the sake of completeness and a general interest, we  use the existing basic method of transformation [1,11] of (5) with (1,4)  to the Gauss hypergeometric equation (GHE) [33] and then we write the solutions for $\psi(x)$ in terms of Jacobi polynomials [33]. The elegant methods  of super-symmetry [7], group-theory [9] and other techniques [13,14] have been used more often in the past. Eventually, we would find familiar expressions wherein it will be interesting  to locate as to how the new interesting results have escaped their discovery earlier [7-11,13,14].

Let us use the transformations [1,11] of $x$ to $z$ and  of $\psi$ to ${\cal F}$ as
\begin{equation}
z=(1-i\sinh x)/2, \quad \psi(z)=  z^{-p}(1-z)^{-q} {\cal F}(z)
\end{equation}
in (5) with (1). After  interesting and involved algebraic manipulations, we get GHE [33] as 
\begin{equation}
z(1-z) {\cal F}''(z)+[-2p+1/2-(-2p-2q+1)z]{\cal F}'(z)-[(p+q)^2+k^2]{\cal F}(z)=0,
\end{equation}
where
\begin{equation}
k=\sqrt{2\mu E}/\hbar, \quad p^2 +p/2 +(P+iQ)/4=0, \quad q^2 +q/2 +(P-iQ)/4=0.
\end{equation}
Using $P$ and $Q$ from (8), two branches of $p$ and $q$ are
\begin{equation}
p=-1/4\pm [1/4+(A+B)/2], q= -1/4\pm [1/4+(A-B)/2].
\end{equation}
Two Linearly independent solutions of  GHE [33] are
\begin{eqnarray}
{\cal F}_1= ~_2F_1[-p-q+ik,-p-q-ik,-2p+1/2;z], \nonumber \\ {\cal F}_2=z^{2p-1/2}~_2F_1[p-q+1/2+ik,p-q+1/2-ik,2p+3/2;z].
\end{eqnarray}
Next, the condition that  bound eigenstates, $\psi(x)$, are $L^2$-integrable requires that $\psi(x)$ must vanish at $x=\pm \infty$. This is met when $-p-q+ik=-n_1$ and $q-p+1/2+ik=-n_2$, separately. These quantization conditions help converting the Gauss hyper geometric functions, ${\cal F}_{1,2}$ (10), to  Jacobi polynomials. We can express two linearly independent  solutions of (5) with (1,4) as
\begin{equation}
\psi_1(x)= {\cal N}_1~ \mbox{sech}^{(p+q)} x ~ \exp[i(p-q)\tan^{-1}(\sinh x)]~ P_{n_1}^{(-2p-1/2,-2q-1/2)}(i\sinh x),
\end{equation}
\begin{equation}
\psi_2(x)= {\cal N}_2~ \mbox{sech}^{(q-p-1/2)} x ~ \exp[-i(p+q+1/2)\tan^{-1}(\sinh x)]~ P_{n_2}^{(2p+1/2,-2q-1/2)}(i\sinh x).
\end{equation}
When $A,B \in {\cal R}$, these two solutions converge asymptotically and  are well known [9,11,13,14] to give rise to two branches of real discrete spectrum and the corresponding eigenstates. These two branches are further distinguished by invoking quasi-parity [10].

This is the point from where we digress from previous works [7-11,13,14] in revealing three parametric  domains of complex, non-Hermitian, non-PT-symmetric Scarf II potentials possessing real discrete spectrum. Firstly,  we allow $A$ to be real but  $B$ remains complex. Secondly, we allow $B$ to be real but $A$ would be complex. It is important to mention that for fixed values of $A$ and $B$ (with only one of them as real)
only one of $\psi_{1,2}(x)$ (11,12) converge asymptotically and is $L^2$-integrable; and the other one diverges. We see these points more precisely in the following three cases. 

\noindent
{\bf Case (1): $A>0$}\\
Taking upper signs in (9) and  using ${\cal F}_1$ (10), we get the usual expressions [8,9] but with a new result that
\begin{equation}
E^A_{n}=-(n-A)^2, ~n=0,1,2,.. < A,~ B=b+i\beta,~ A, ~ b,~ \beta \in {\cal R}.
\end{equation}
The solution (11) becomes
\begin{equation}
\psi^A_{n}(x)= {\cal N}_A ~(\mbox{sech}x)^A ~ \exp[iB \tan^{-1}(\sinh x)]~ P_{n}^{(-A-B-1/2,-A+B-1/2)}(i\sinh x).
\end{equation}

\noindent
The asymptotic convergence of the other linearly independent solution (see Eq. (17) below)  cannot be controlled by the condition $n < A$ (Eq. (13)). Even if
$\psi_2(x)$ converges asymptotically $n (=n_2)$ would no more be real as $p-q+1/2+ik=-n_2 \Rightarrow B+1/2=-n_2$ and $B$ has been  deliberately chosen to be complex.

\noindent
{\bf Case (2): $A<0$}\\
Taking lower signs in (9) and 
using ${\cal F}_2$ (10), we get 
\begin{equation}
E^{-A}_{n}=-(n-(-A-1))^2, ~n=0,1,2,.. < -A-1,~ B=b+i\beta,~ A, ~ b,~ \beta \in {\cal R}.
\end{equation}
It can be readily checked that when in (1,4) $A<0$ the depth of the real part becomes less (shallow well) so it has  lesser number of bound states as compared to the case when $A>0$ in (1,4). Earlier, in the literature [2-5,7-11,13,14] this case has been unnoticed for both Hermitian and complex PT-symmetric Scarf II. Using the solution (11) again, we get the corresponding eigenstate 
\begin{equation}
\psi^{-A}_{n}(x)= {\cal N}_A ~(\mbox{sech}x)^{-A-1}~ \exp[-iB \tan^{-1}(\sinh x)]~ P_{n}^{(A+B-1/2,A-B-1/2)}(i\sinh x).
\end{equation}

\noindent
{\bf Case (3): $B \in {\cal R}$}\\ 
It can also be checked that when $B$ is real, both options:  upper and  lower signs in Eq. (9) degenerate to one case (as opposed to the Cases (1,2) when $A \in {\cal R}$ in above). Physically, this reflects the invariance of eigenvalues  under reflection of Scarf II (1,4) : $V(-x,-B)=V(x,B)$.
Using ${\cal F}_2$ in (10), we again get usual expressions but another new result that
\begin{equation}
E^{|B|}_{n}=-(n-|B| + 1/2)^2,~ n=0,1,2,3,..< |B|-1/2,~ A=a+i\alpha,~ B,~ a,~\alpha \in {\cal R};
\end{equation}
\begin{equation}
\psi^{|B|}_{n}(x)={\cal N}_{|B|} ~ (\mbox{sech} x)^{|B|-1/2} ~ \exp[-i(A+1/2) \tan^{-1} (\pm \sinh x)]~ P_{n}^{(A+|B|+1/2,-A+|B|-1/2)} (\pm i\sinh x),
\end{equation}
where the upper (lower) signs are for $B>0(<0)$. 
In Eqs. (14,16,18), notice that the argument of exponential  remains finite for any value of $x$ irrespective of whether  $A$ or $B$ are  positive/negative or non-real. Similarly, the functions  $P^{(c,d)}_n(i \sinh x)$  remain  polynomials of $(i\sinh x)$ of degree $n \in I^++\{0\}$ 
with real or non-real coefficients. Eventually, the presence of  $\mbox{sech}^{\nu}x, \nu >0$ for the allowed values of $n$ dampens $\psi_n(x)$, asymptotically on both sides. These three aspects of  $\psi_n(x)$ ensure $L^2$-integrability of the eigenstates in all three cases discussed above.

In all the calculations presented here, we assume $2\mu=1=\hbar^2$. In Fig. 1, we show the complex non-PT-symmetric potential (1,4) for $A=2.7, B=1.2+1.4i$ and the ground state complex eigenfunction (14). This potential
possesses three real discrete eigenvalues: $E_0= -7.29, E_1=-2.89, E_2=-0.49$, see Eq.(13). In Fig. 2, we show complex non-PT-symmetric Scarf II (1,4) for $A=-2.7, B=1.2 +1.4i$ and its ground state complex eigenfunction, see Eq. (16). There are only two real discrete eigenvalues at $E_0=-2.89$ and  $E_1=-0.49$, see Eq.(15). In Fig. 3, we show the potential (1,4) and the ground state eigenfunction (18) for $A=-2.3 + 1.1 i, B= 3.1$. This potential has three real bound states with eigenvalues: $E_0=-6.76, E_1= -2.56$ and  $E_2=-0.36$, see Eq.(17). Not shown here are the other (higher) eigenstates of these potentials, all of them like the eigenstates in Figs. 1-3 are $L^2$-integrable. 
In Figs. 1-3, one may observe that at least the real part of the new  potentials constitute a prominent well. For a given set of parameters as taken in Figs. 1-3, we have checked that the various eigenstates of a fixed potential are orthogonal as
\begin{equation}
\int_{-\infty}^{\infty} \psi_{m} (x) \psi_{n}(x) dx=0,\quad E_m \ne E_n.
\end{equation}
This orthogonality is due to the $L^2$-integrability of the new eigenstates which are finite everywhere in $(-\infty,\infty)$ and follow Neumann boundary condition that $\psi(\pm \infty)=0$. See the Appendix for a proof.

In Hermitian quantum mechanics the physical poles of $r(k)$ and $t(k)$ yield the possible real discrete spectra of the potential. More interestingly, the negative energy (bound-state) eigenvalues  manifest as poles of the transmission, $T(E)=|t|^2$, and reflection, $R(E)=|r|^2$, coefficients. See Fig. 3(a)
for the location of negative energy bound state eigenvalues of a real square-well potential (of depth $|V_0|=5$ and width $4$ units)  as the negative-poles in $T(E)$ and $R(E)$. The deep minima (zeros) in $R(E)$ indicate the presence of  anti-bound states. We find that the scattering coefficients of complex PT-symmetric potentials also share this property  in yielding [23] the possible real discrete  bound states of  a complex scattering potential. Let us remark that the negative-energy  poles of $R_{left}(E), R_{right}(E), T(E)$, despite the non-reciprocity of the reflection coefficient, $R_{left}(E) \ne R_{right}(E)$ [16], coincide at the bound states of a complex PT-symmetric potential, see Fig. 3(b). Since for a complex scattering potential $T(E)$ is invariant [16] of the direction of incidence i.e.,$ T_{left}(E)=T_{right}(E)$ [16],
we can expect the negative-energy poles of $T(E)$ to yield  the possible discrete spectrum in an unambiguous way. It is intriguing to see that as a complex potential deviates significantly from a PT-symmetric potential through the real parameters, $\alpha$ and $\beta$ in Eqs. (13,15,17),  the negative poles of  all three coefficients do not coincide. For the newly proposed non-PT-symmetric potentials with $n$ number of real discrete eigenvalues, we see (Figs. 5-7) a new scenario wherein  $n$ number of negative energy poles of $T(E)$ are found to coincide with  $n$  number of poles of either  $R_{left}(E)$ or $R_{right}(E)$. 

In Figs. 5-7, we calculate the $T(E)$ (solid line) and $R(E)$ (dashed line) for the Scarf potential (1,4) using the analytic expressions of $r(k)$ and $t(k)$ as given in [5,15]. A small correction to $t(k)$ as proposed in [15] consists of replacing $\Gamma(1+ik)$ in (Eq. (18a) of Ref. [5])  by $\Gamma(1-ik)$ [15]. We confirm  this  and find that it  matters specially when we  study $r(k)$ and $t(k)$ at negative energies for bound states of the potential (1,4) for various cases as given in Figs. 1-3. 

For the proposed complex non-PT-symmetric Scarf II, the parts (a) in Figs. 4-7  display $R_{left}(E)$ and (b) display $R_{right}(E)$ by dashed lines. Notice that both the have common and uncommon poles but only 2/3  of them coincide with the poles of $T(E)$ (solid lines) which are 2/3 real discrete eigenvalues of the newly proposed non-PT-symmetric domains of complex Scarf II potential. Not shown here are $T(E)$ and $R(E)$ for positive energies, we find
an absence of spectral singularity [17,22,24] in them. One usually finds [26] that bound states and spectral singularities  are mutually exclusive in a  fixed complex  potential. However, it requires a proof either ways.

Thus we find that even the most common quantum Hamiltonians $H=p^2/2\mu + V (x)$ are able to generate the real bound-state-simulating spectra {\em without} the (in a way, redundant) assumption of their parity-times-time-reversal symmetry {\em alias} parity-pseudo-Hermiticity.
Also the absence of any metric ($\eta$) in the orthogonality condition (19) seems to deny the pseudo-Hermiticity of $H$ in the proposed new parametric domains of complex Scarf II.

In phenomenological context, paradoxically enough, this result could
prove also (or even more) interesting in classical optics these
days. Indeed, once one translates the quantum-mechanical concept of
PT symmetry in the most contemporary and experiment-oriented
simulation terminology of loss-gain-systems in paraxial
approximation, one could also appreciate the second half of the message, viz., the study of scattering amplitudes and of their singularities.

We have established the presence of three  non-PT-symmetric parametric regimes in complex Scarf II potential where it possesses real discrete spectra and the energy eigenstates 
are orthogonal with regard to a simple scalar product.
Given the extensive investigations of the complex Scarf II potential for more than one and a half decade this result is an interesting surprise. When a complex PT or non-PT-symmetric Hamiltonian has real discrete spectrum it is also found to be $\eta$-pseudo-Hermitian under one or more metrics ($\eta$), we feel that it will be challenging to show the Hamiltonian for  non-PT-symmetric branches of  complex Scarf II(proposed here) as pseudo-Hermitian, if possible. The features  of the negative energy poles of reflection/transmission  amplitudes/coefficients of complex non-PT-symmetric potential are new and they open a scope for  further investigations in both the classical and quantum optics.  
\begin{figure}[H]
\centering
\includegraphics[width=7 cm,height=5. cm]{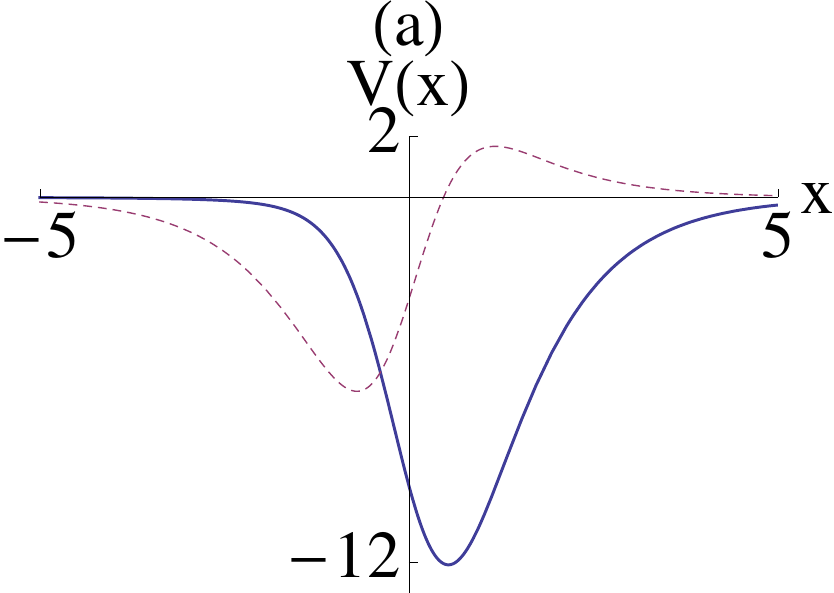}
\hskip .5 cm
\includegraphics[width=7 cm,height=5. cm]{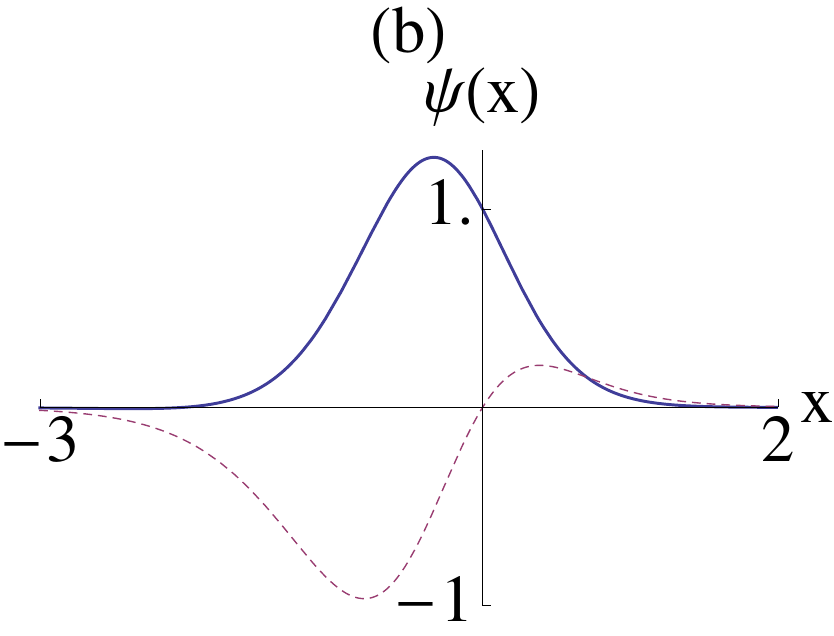}
\caption{(a) Real (solid line) and imaginary (dashed line) parts of the complex non-PT-symmetric potential, $V(x)=(-9.47-3.36i)~\mbox{sech}^2x+(-8.96+7.68i) \tanh x~ \mbox{sech} x$, arising from Eqs. (1,4) in Case (1): $A=2.7, B=1.2+1.4i$. It has three real discrete eigenvalues as $E_0=-7.29, E_1=-2.89, E_2=-0.49$. (b) Shows real (solid line) and imaginary (dashed line) parts of the ground eigenstate.}
\end{figure}
\begin{figure}[H]
\centering
\includegraphics[width=7 cm,height=5. cm]{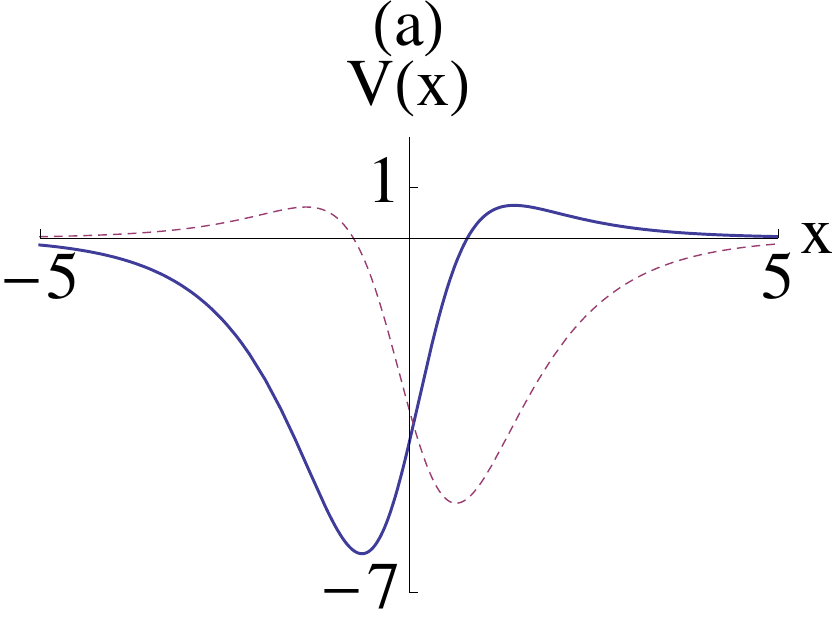}
\hskip .5 cm
\includegraphics[width=7 cm,height=5. cm]{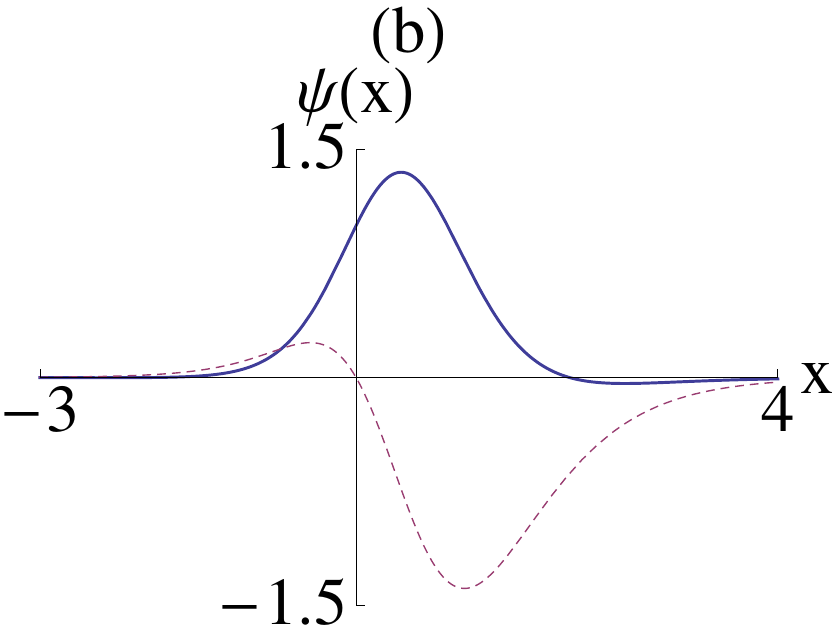}
\caption{ The same as in Fig. 1, for $V(x)= (-4.07-3.36i)~ \mbox{sech}^2 x +(6.16-5.28i)\tanh x~ \mbox{sech}x$ in Case (2) when $A=-2.7, B=1.2+1.4i$. Here $E_0=-2.89$ and $E_1=-0.49.$}
\end{figure}
\begin{figure}[H]
\centering
\includegraphics[width=7 cm,height=5. cm]{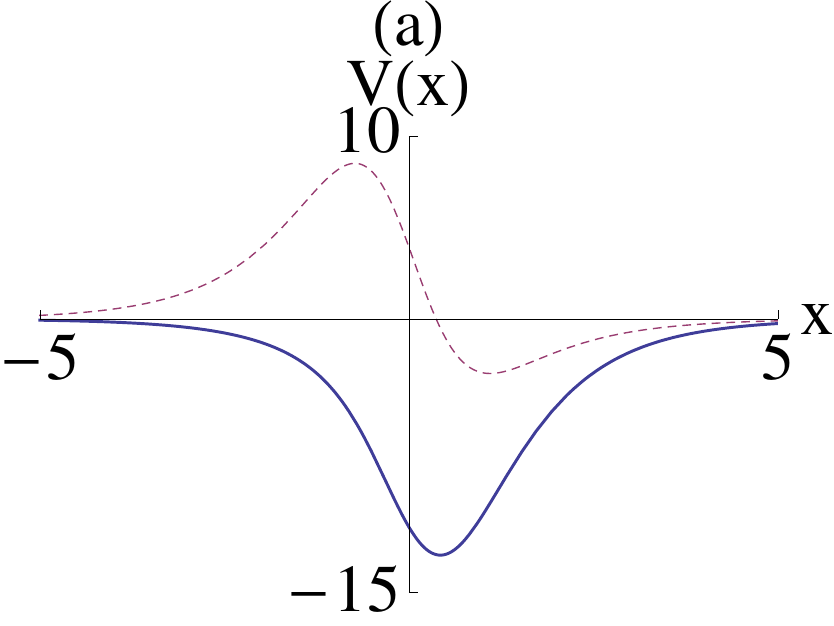}
\hskip .5 cm
\includegraphics[width=7 cm,height=5. cm]{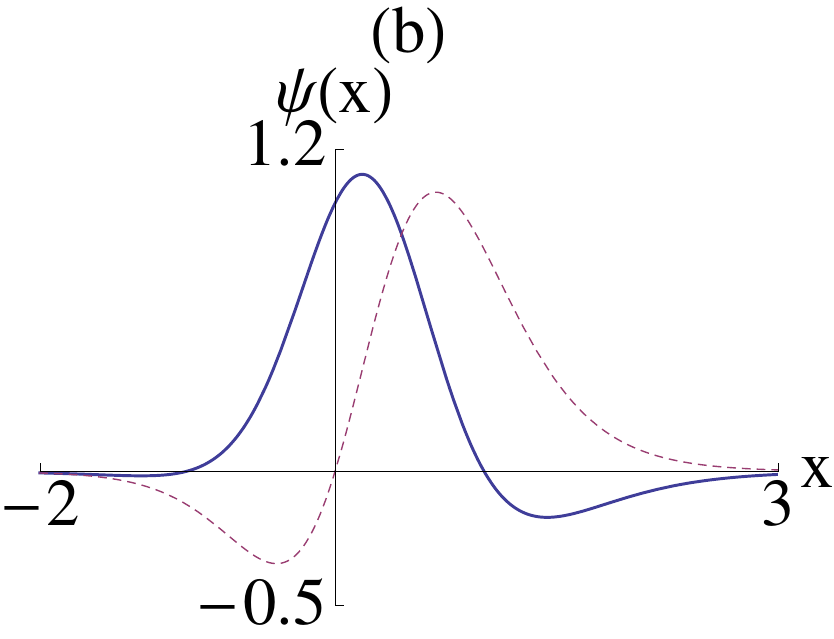}
\caption{The same as in Fig. 1, for $V(x)=(-11.39+3.96)~ \mbox{sech}^2 x-(6.82+11.16i) \tanh x~ \mbox{sech} x $ in Case (3) when $A=-2.3+1.1i, B=\pm 3.1$. The eigenvalues are $E_0=-6.76, E_1=-2.56$ and $E_0=-0.36.$}
\end{figure}
\begin{figure}[H]
\centering
\includegraphics[width=7 cm,height=5. cm]{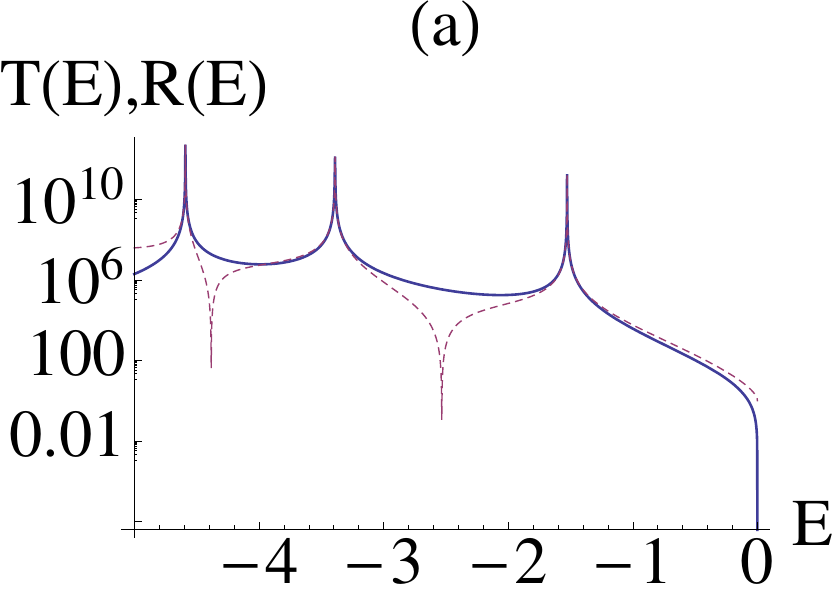}
\hskip .5 cm
\includegraphics[width=7 cm,height=5. cm]{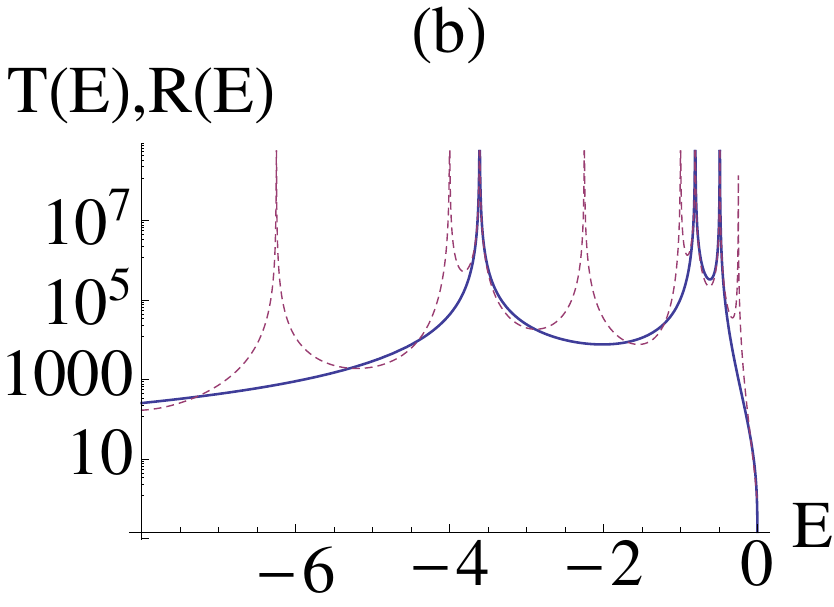}
\caption{(a) $T(E)$ (solid line) and $R(E)$ (dashed line) for the usual Hermitian square-well potential of depth 5 and width 4, its eigenvalues are $E_0=-4.59, -3.38, -1.52$. Notice the coincidence of the poles in both at these energies. The deep minima (zeros) in $R(E)$ indicate the anti-bound states. (b) $T(E)$ (solid line) and  $R_{left}(E)$, $R_{right}(E)$ (coincident dashed lines) for complex PT-symmetric Scarf II with $A=1.9, B=1.2$. Notice the coincidence of the poles of $T(E)$ and $R(E)$ at
$ E_0=-3.61, E_1=-0.81$ and $E_0=-0.49$ (second branch).}
\end{figure}
\begin{figure}[H]
\centering
\includegraphics[width=7 cm,height=5. cm]{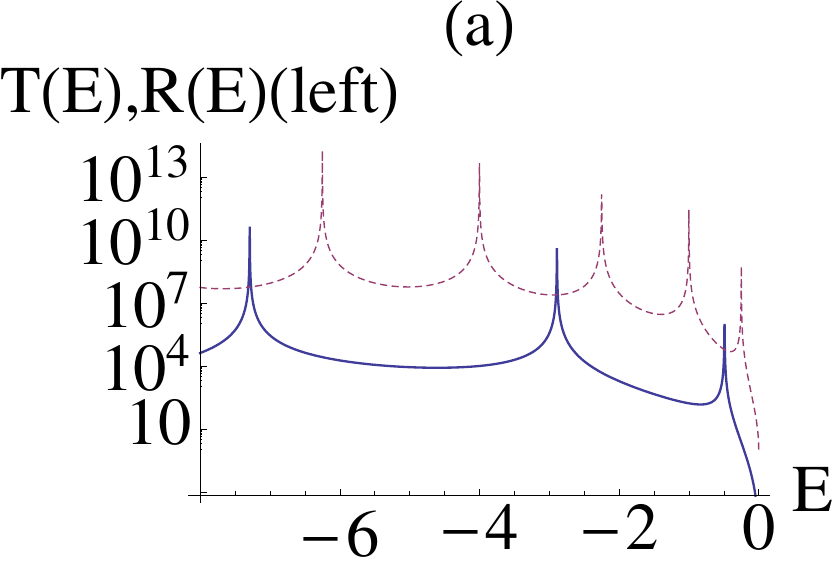}
\hskip .5 cm
\includegraphics[width=7 cm,height=5. cm]{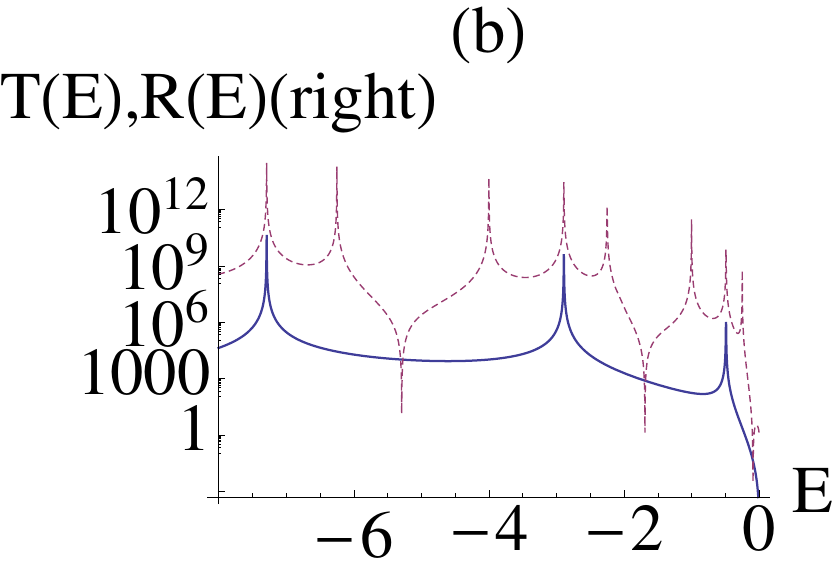}
\caption{The same as in Fig. 4 for $V(x)$ of Fig. 1. Notice
the coincidence of poles of $T(E)$ and $R_{right}$ at the boundstate eigenvalues:  $E_0=-7.29, E_1=-2.89, E_2=-0.49$. Also notice that $R_{left}(E)$ and $R_{right}(E)$ have both common and uncommon poles.}
\end{figure}
\begin{figure}[H]
\centering
\includegraphics[width=7 cm,height=5. cm]{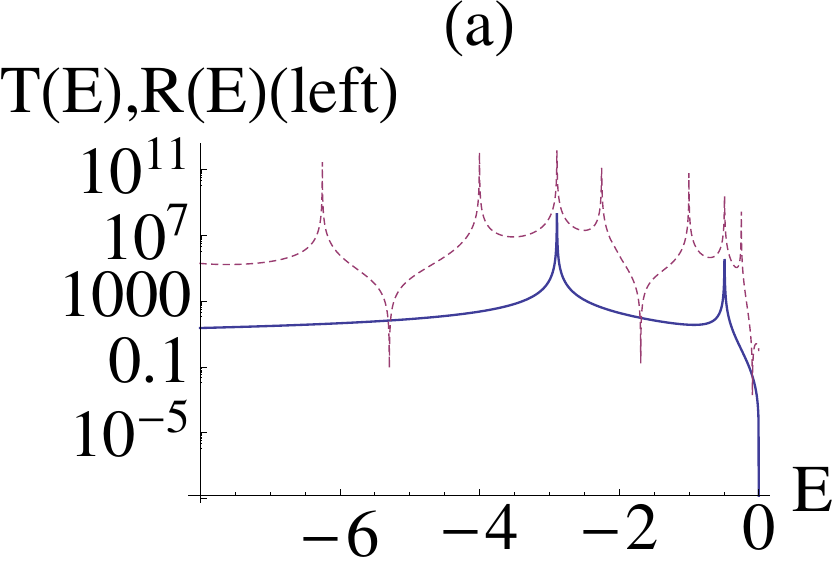}
\hskip .5 cm
\includegraphics[width=7 cm,height=5. cm]{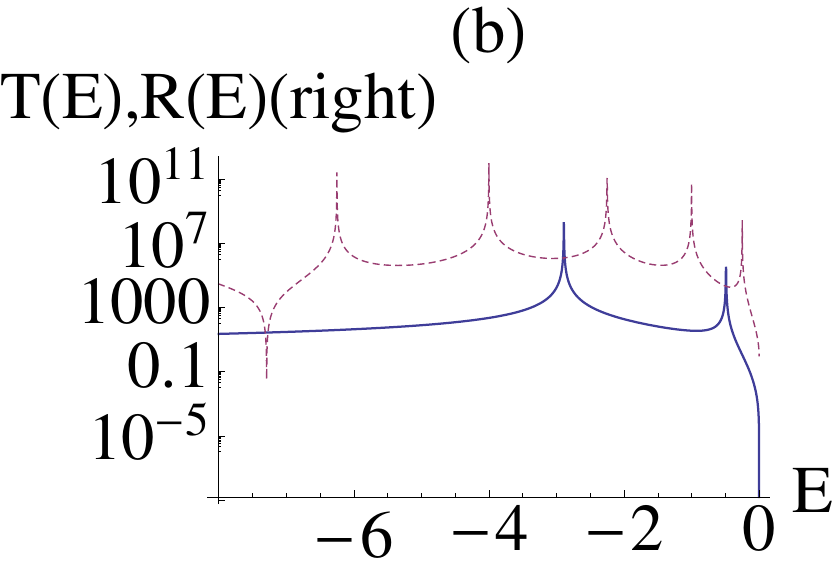}
\caption{The same as in Figs (4,5) for the potential in Fig. 3. The poles of $T(E)$ and $R_{left}$ are coincident at $E_0=-2.89, E_1=-0.49$.}
\end{figure}
\begin{figure}[H]
\centering
\includegraphics[width=7 cm,height=5. cm]{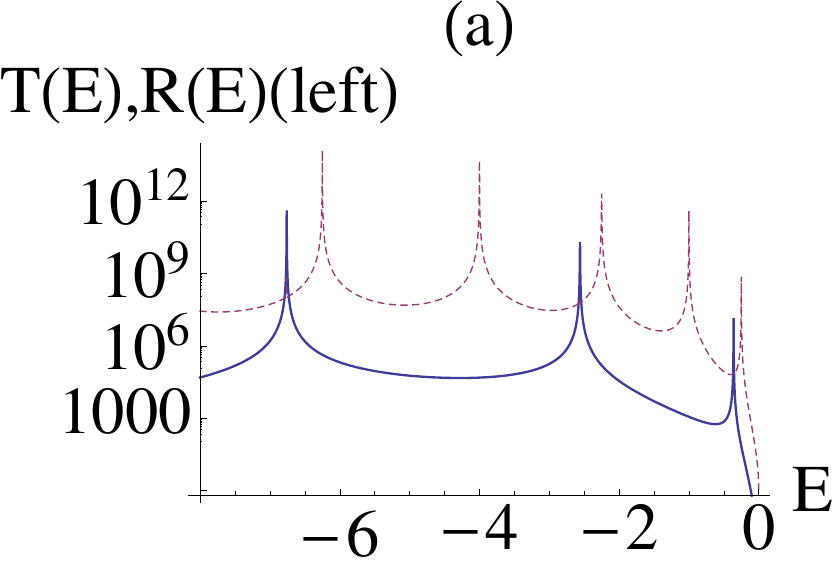}
\hskip .5 cm
\includegraphics[width=7 cm,height=5. cm]{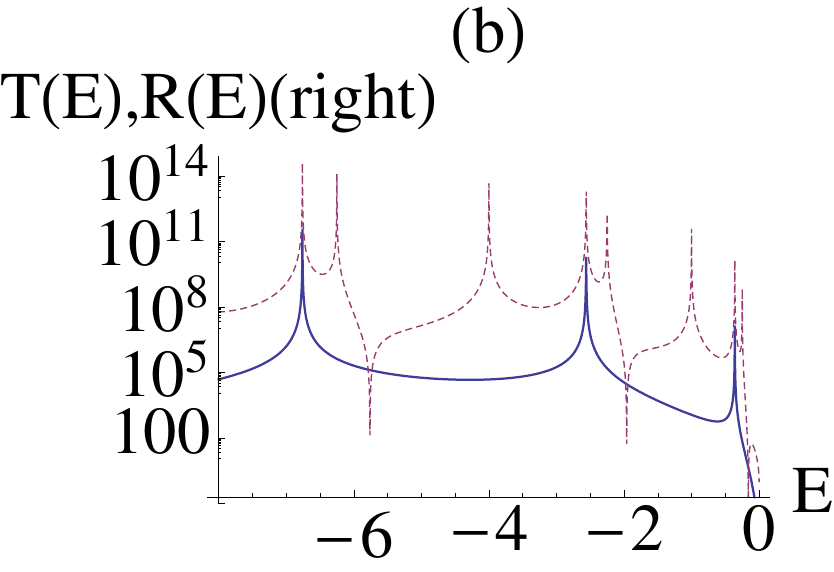}
\caption{The same as in Figs. 4,5 for the potential in Fig. 3. This time poles of $T(E)$ and $R_{right}(E)$ are coincident at $E_0=-6.76, E_1=-2.56, E_0=-0.36$.}
\end{figure}
\renewcommand{\theequation}{A-\arabic{equation}}
\setcounter{equation}{0}
\section*{Appendix}
Having discussed that three types (domains) of complex
non-Hermitian Scarf II potentials possess
the $L^2$-integrable eigenstates which are finite
for $x \in (-\infty, \infty)$  and satisfy Neumann condition that $\psi(\pm \infty) =0$, we now prove their 
orthogonality as proposed by us in Eq. (19).
Let $\psi_m(x), \psi_n(x)$ be two solutions of Schr{\"o}dinger equation (5) with distinct real eigenvalues $E_m$ and $E_n$, then we have
\label{allequations}
\begin{eqnarray}
\frac{d^2\psi_m(x)}{dx^2}+[E_m-V(x)]\psi_m(x)=0, \\ 
\frac{d^2\psi_n(x)}{dx^2}+[E_n-V(x)]\psi_n(x)=0.
\end{eqnarray}
Multiply the first by $\psi_n(x)$ and the second by $\psi_m(x)$  and by subtracting them we get
\begin{equation}
(E_m-E_n) \psi_m(x) \psi_n(x) =\psi_m(x) \frac{d^2\psi_n(x)}{dx^2}-\psi_n(x) \frac{d^2\psi_m(x)}{dx^2}= \frac{d}{dx}\left (\psi_m \frac{d\psi_n}{dx}-
\psi_n \frac{d\psi_m}{dx}\right).
\end{equation}
Integrating the above equation from $x=-\infty$
to $x=\infty$, we get 
\begin{equation}
(E_m-E_n)\int_{-\infty}^{\infty} \psi_m(x) \psi_n(x)~dx= 
\left (\psi_m(x) \frac{d\psi_n}{dx}-
\psi_n(x) \frac{d\psi_m}{dx} \right)_{-\infty}^{\infty}.
\end{equation}
Since $\psi(\pm\infty)=0$, we finally prove (19)
\begin{equation}
\int_{-\infty}^{\infty} \psi_m(x) \psi_n(x)~dx=0, \quad E_m \ne E_n.
\end{equation}

\end{document}